\documentclass[review]{elsarticle}

\usepackage{hyperref}
\usepackage{amsmath}
\usepackage{geometry}
\usepackage{setspace}

\journal{Icarus}

\usepackage{numcompress}\biboptions{authoryear}

\begin{document}

\begin{frontmatter}

\title{Investigating the Semiannual Oscillation on Mars using data assimilation}

\author[oxford]{Tao Ruan}
\author[oxford]{Neil T. Lewis}
\author[ou]{Stephen R. Lewis}
\author[oxford,ssi]{Luca Montabone}
\author[oxford]{Peter L. Read\corref{correspondence}}
\cortext[correspondence]{Corresponding author: Peter Read, 
                         Atmospheric, Oceanic and Planetary Phyiscs, 
                         Clarendon Laboratory, Oxford, OX1 3PU, UK. 
                         Tel: +44 1865 272082.
                         }
\ead{peter.read@physics.ox.ac.uk}

\address[oxford]{Atmospheric, Oceanic and Planetary Physics, Clarendon Laboratory, Oxford, OX1 3PU, UK}
\address[ou]{School of Physical Sciences, The Open University, Walton Hall, Milton Keynes, MK7 6AA, UK}
\address[ssi]{Space Science Institute, Boulder, CO, USA}

\begin{abstract}
A Martian semiannual oscillation (SAO), similar to that in the Earth’s tropical stratosphere, is evident in the Mars Analysis Correction Data Assimilation reanalysis dataset (MACDA) version 1.0, not only in the tropics, but also extending to higher latitudes. Unlike on Earth, the Martian SAO is found not always to reverse its zonal wind direction, but only manifests itself as a deceleration of the dominant wind at certain pressure levels and latitudes. Singular System Analysis (SSA) is further applied on the zonal-mean zonal wind in different latitude bands to reveal the characteristics of SAO phenomena at different latitudes. The second pair of principal components (PCs) is usually dominated by a SAO signal, though the SAO signal can be strong enough to manifest itself also in the first pair of PCs. An analysis of terms in the Transformed Eulerian Mean equation (TEM) is applied in the tropics to further elucidate the forcing processes driving the tendency of the zonal-mean zonal wind. The zonal-mean meridional advection is found to correlate strongly with the observed oscillations of zonal-mean zonal wind, and supplies the majority of the westward (retrograde) forcing in the SAO cycle. The forcing due to various non-zonal waves supplies forcing to the zonal-mean zonal wind that is nearly the opposite of the forcing due to meridional advection above ${\sim}3$ Pa altitude, but it also partly supports the SAO between 40 Pa and 3 Pa. Some distinctive features occurring during the period of the Mars year (MY) 25 global-scale dust storm (GDS) are also notable in our diagnostic results with substantially stronger values of eastward and westward momentum in the second half of MY 25 and stronger forcing due to vertical advection, transient waves and thermal tides.
\end{abstract}


\end{frontmatter}

\section{Introduction}

In the upper stratosphere and mesosphere of Earth, the semiannual oscillation (SAO) of the mean zonal wind in the tropical stratosphere and mesosphere is commonly observed as a clear feature \citep{1966JGR....71.4223R,1997JGR...10226019G}. \citet{1966JGR....71.4223R} used observational data to determine that the strongest westerly winds occurred shortly after the equinoxes in the lower mesosphere and then later extended to the lower levels. The pronounced SAO occurs at intermediate and upper levels of the stratosphere and mesosphere, and reaches a peak in amplitude at about the height of the stratopause. 
  
Similar SAO-like features in the Martian tropics (between $10^{\circ}\text{S}$ - $10^{\circ}\text{N}$) have been noted in the work of \citet{2008GeoRL..3523202K}, mainly based on a free-running Martian Global Climate Model (MGCM). Suspended dust particles are widespread in the Martian atmosphere, and they strongly affect its dynamical and thermal structure. Studying and characterizing the SAO phenomenon on Mars may contribute to understanding the dust cycle on Mars, as well as the Martian weather and climate. 

So far, no comprehensive measurements of wind fields are available for the Martian atmosphere, although some localized snapshot measurements of middle atmosphere winds have been obtained from ground-based microwave measurements \citep{2012Icar..217..315S}. At middle and high latitudes, horizontal winds may be approximately computed using the thermal wind relation assuming geostrophic or gradient wind balance, via temperature fields observed by the instruments on board various satellites (e.g., Mars Global Surveyor (MGS) - \citealp{2000JGR...105.9509C}, \citealp{2008AREPS..36..191S}; and Mars Climate Sounder (MCS) - \citealp{2010JGRE..11512016M}). However, even the gradient thermal wind relation becomes invalid at low latitudes so tropical winds are difficult to recover. 

On the other hand, from variations in the structure and amplitude of solar thermal tides it may be possible to determine indirectly the corresponding variations of zonal-mean zonal wind. On Earth, \citet{1974JAtS...31.1421L} discovered that the zonal wind could affect the propagation of solar tides. \citet{1996JAtS...53.1290W} also showed evidence of the effect of the solstitial mean zonal wind on the tide propagation on Mars. The seasonally varying wind should therefore also affect the pattern of solar tides. In the work of \citet{2008GeoRL..3523202K}, the SAO signal was evident by analyzing the solar thermal tides, derived directly from the observed temperature fields (from the Thermal Emission Spectrometer - TES onboard MGS) by taking the difference between the daytime and nighttime thermal fields. The thermal tides showed a clear SAO variation between 50 and 5 Pa in their study. 
  
To conduct their detailed study of the SAO, \citet{2008GeoRL..3523202K} made use of a simulation from a free-running MGCM. A clear SAO phenomenon between $10^{\circ}\text{S}$ - $10^{\circ}\text{N}$, similar in form to the stratospheric SAO on Earth, was described in their results \citep[Figure 2 in][]{2008GeoRL..3523202K}. To investigate the driving mechanism of the Martian SAO within the tropics, \citet{2008GeoRL..3523202K} went on to calculate the different contributions to the tendency of the mean zonal wind, $\partial\overline{u}/\partial t$ on interseasonal timescales. Their conclusion was that, at solstices, the horizontal cross-equatorial transport provided the westward forcing to the tropical zonal-mean zonal wind and vertical advection mostly provided a source of eastward momentum. The solar tides supplied eastward forcing at equinoxes. Somewhat surprisingly, at all seasons, Kelvin waves appeared to provide a westward forcing at all heights. Unlike on Earth, however, the effects of transient planetary waves and the eastward Kelvin waves on the Martian SAO were both quite weak.

Since the SAO may have further implications for the transport of tracers on Mars, as it has in the Earth’s stratosphere, it may be important to investigate the phenomenon of the Martian SAO in detail, with the additional aim of understanding the Martian dust cycle more completely. The intermittent nature of the available satellite measurements requires an interpolation in terms of time and space to acquire a complete observational coverage for the diagnosis of the Martian SAO. On the other hand, numerical models are able to overcome this difficulty to provide consistent data in both time and in three-dimensional space. A MGCM is therefore a perfect tool for analyzing the Martian SAO. The modeling study of \citet{2008GeoRL..3523202K} seemed to provide a reasonably plausible representation of the Martian SAO in the tropics. But  is a free-running MGCM, run with an unrealistically uniform dust distribution, sufficient to represent this aspect of the real Mars atmospheric environment? Various numerical modeling studies have proved capable of reproducing the full seasonal variability of the Martian climate, but MGCMs still struggle to fully represent the realistic interannual variability of the Martian climate system \citep{2002JGRE..107.5123N,2002JGRE..107.5124N,2004JGRE..10911006B,2006JGRE..111.9004B,2012..Thesis..M,2013Icar..223..344M}. In this context, data assimilation offers another approach towards obtaining a four-dimensional representation of atmospheric behavior that has the additional advantage of being consistent with both Martian observations and modeled physical constraints \citep{1995AdSpR..16....9L,2006Icar..185..113M,2007Icar..192..327L}. 

The goal of the present study is to diagnose the Martian SAO, based on the results from a Martian data assimilation system that is tightly constrained by real observations and can therefore provide a more realistic climatology than a free-running MGCM. The phenomenology and driving forces of the observed Martian SAO will be presented and investigated in this study. We wish to confirm if we observe a similar phenomenon in the tropics of the assimilated Martian climate measurements to that found in the work of \citet{2008GeoRL..3523202K}, and if the processes identified as driving the observed SAO are consistent with their analysis. A further question that can be answered by this observationally-based approach is whether components of an SAO can be observed outside of the Martian tropics.

\section{Data and methodology}

The dataset used in this study for diagnosing the Martian SAO is the MGS-TES Mars Analysis Correction Data Assimilation (MACDA) version 1.0 \citep{2011..MACDA,2014GSDJ....1..129M}. MACDA v1.0 is based on the UK-LMD MGCM that was first described in the work of \citet{1999JGR...10424155F}. A successive analysis correction scheme \citep{1991QJRMS.117...59L} was adapted and developed to assimilate retrieved temperature profiles and column-integrated dust opacities into the UK-LMD MGCM \citep{2007Icar..192..327L}. Further technical details of this reanalysis database can be found in the work of \citet{2014GSDJ....1..129M}.  

MACDA v1.0 covers nearly three complete Mars years (MYs), corresponding to the MGS-TES observational periods from $141^{\circ}$ solar longitude ($L_{s}$) in MY 24 through to $L_{s} = 82^{\circ}$ in MY 27. \citet{2006Icar..185..113M} and \citet{2007Icar..192..327L} have already validated this dataset against available observations (other MGS-TES and radio occultation data), and this dataset has been proven to be reasonably consistent with those observations throughout this period. 

MACDA is provided at a standard resolution equivalent to horizontal spherical harmonic spectral truncation T31, corresponding to a $3.75^{\circ}\times3.75^{\circ}$ nonaliasing dynamical grid. For this analysis, the Martian atmosphere was divided into 7 latitude bands ($60^{\circ}$ - $90^{\circ}$N, $40^{\circ}$ - $60^{\circ}$N, $10^{\circ}$ - $40^{\circ}$N, $10^{\circ}$N - $10^{\circ}$S, $10^{\circ}$ - $40^{\circ}$S, $40^{\circ}$ - $60^{\circ}$S, $60^{\circ}$ - $90^{\circ}$S) for investigating the Martian SAO, both within and outside of the tropics. In section \ref{sec:phenomenology}, the actual zonal wind from MACDA is shown to demonstrate the presence of a Martian SAO. Singular Spectrum Analysis (SSA) \citep{1984...pikegeneralised} is further used in each latitude band to reveal and decompose the phenomenology of the Martian SAO. In section \ref{sec:tem}, the Transformed Eulerian Mean equations (TEM) \citep{1978JAtS...35..175A} are applied to study the different contributions to the tendency of the zonally-averaged zonal wind in the tropical middle atmosphere throughout the seasonal cycle. The pressure and pseudo-height values used in the following figures were simply calculated from the model terrain-following vertical sigma coordinate, using a reference surface pressure of 610 Pa and a uniform assumed scale height of 10 km. It is noted that, because TES data only extends to ${\sim}10$ Pa (i.e. ${\sim}40$ km altitude), the reanalysis above that level is only indirectly influenced by observational increments being made below. Also, as the TES nadir data is only provided at the local time ${\sim}$02:00 and ${\sim}$14:00, this might not fully resolve the tides. Therefore, the reader should be cautioned that the tides in the reanalysis at altitudes above 10 Pa might mainly rely on the UK-LMD MGCM itself rather than on the data assimilated.

\section{Revealing the phenomenology of the observed SAO}\label{sec:phenomenology}

\subsection{Zonal wind}

\begin{figure}
    \centering
    \includegraphics[scale=0.65]{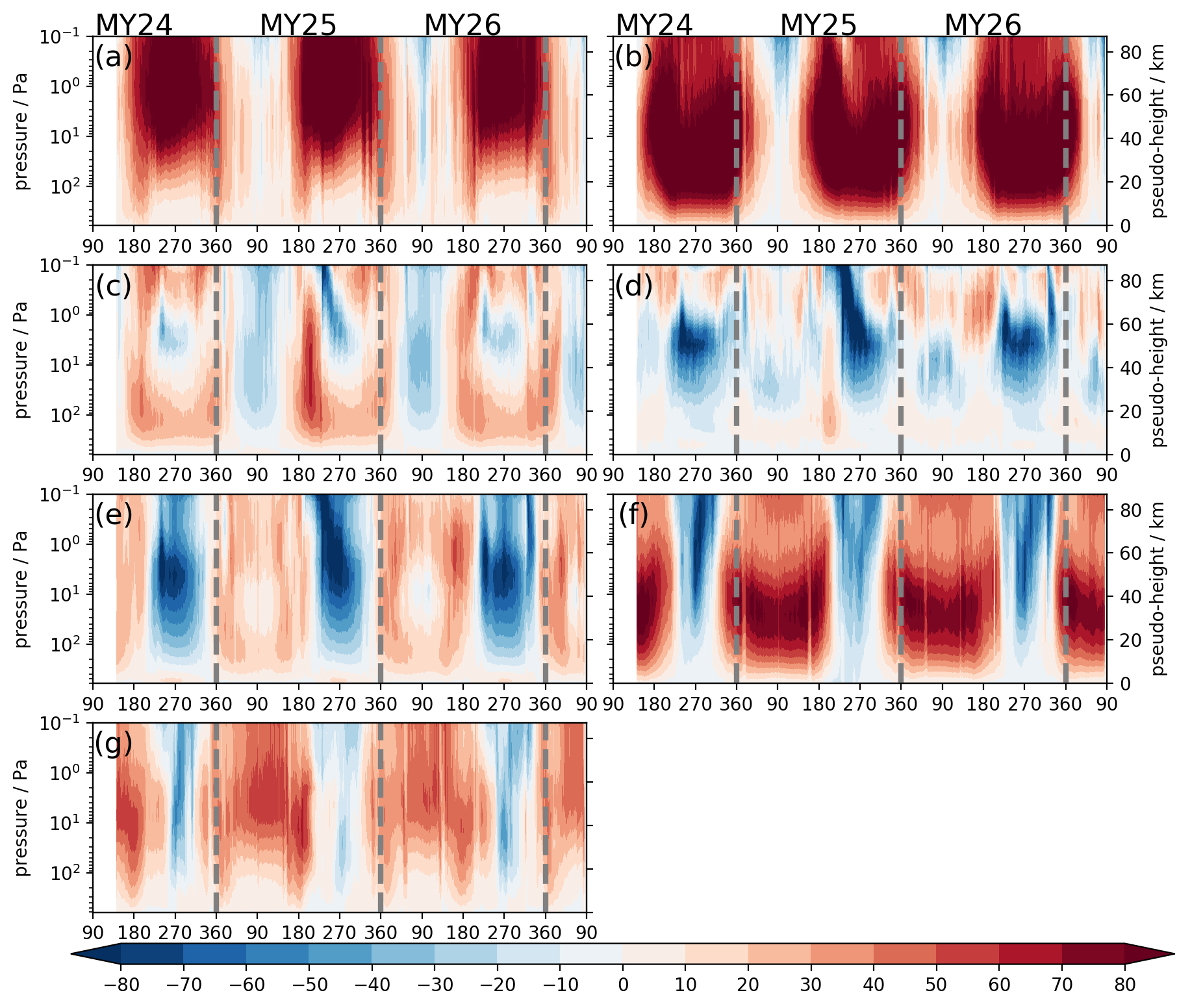}
    \caption{The zonal-mean of daily-averaged MACDA zonal winds (m\,s$^{-1}$) in different latitude bands, (a) $60^{\circ}$ - $90^{\circ}$N, (b) $40^{\circ}$ - $60^{\circ}$N, (c) $10^{\circ}$ - $40^{\circ}$N, (d) $10^{\circ}$N - $10^{\circ}$S, (e) $10^{\circ}$ - $40^{\circ}$S, (f) $40^{\circ}$ - $60^{\circ}$S, (g) $60^{\circ}$ - $90^{\circ}$S.}\label{fig:one}
\end{figure}

\begin{figure}
    \centering
    \includegraphics[scale=0.65]{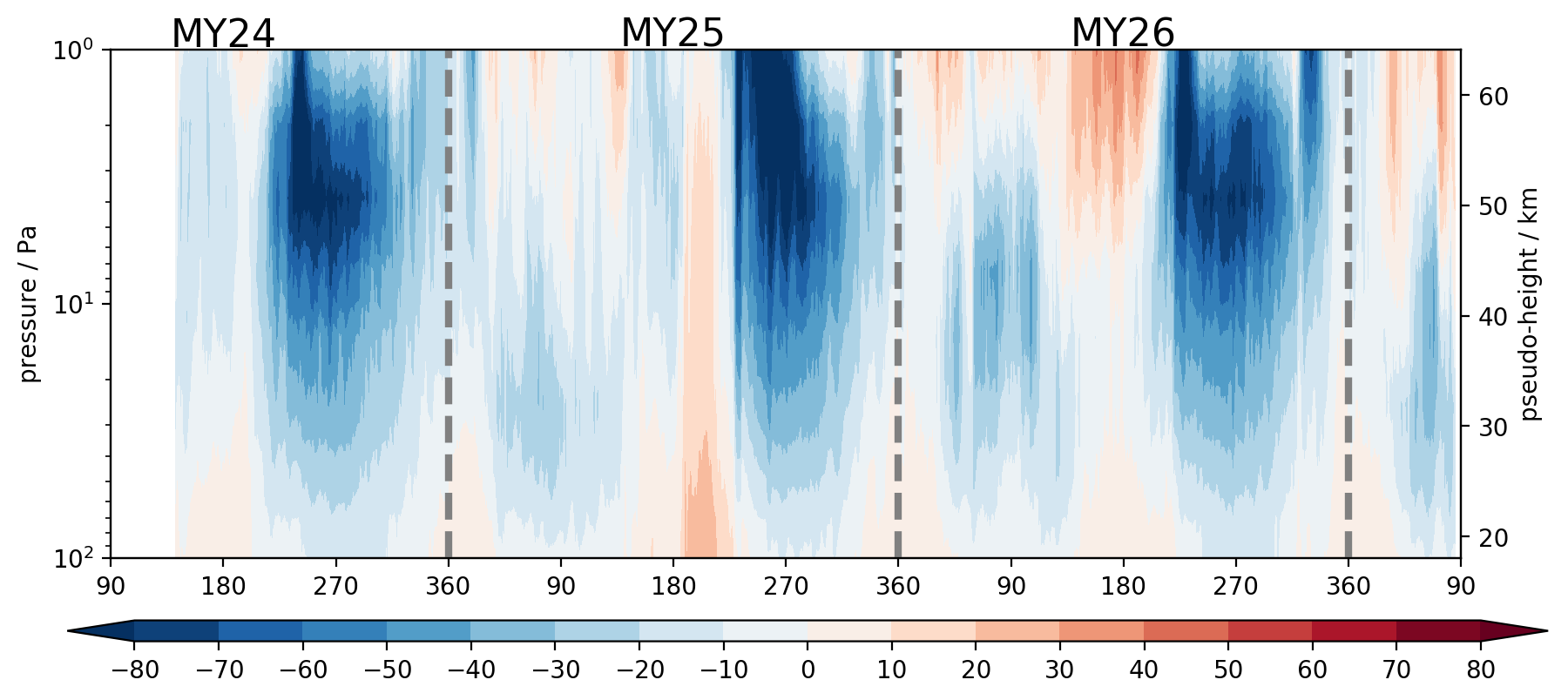}
    \caption{The zonal-mean of daily-averaged MACDA zonal winds (m\,s$^{-1}$) at $10^{\circ}$N - $10^{\circ}$S and between 100 and 1 Pa.}\label{fig:two}
\end{figure}

Figure \ref{fig:one} presents the zonal mean of the daily-averaged zonal wind from reanalysis data (nearly 3 MYs in total) derived from MACDA in the different latitude bands defined above ($60^{\circ}$ - $90^{\circ}$N, $40^{\circ}$ - $60^{\circ}$N, $10^{\circ}$ - $40^{\circ}$N, $10^{\circ}$N - $10^{\circ}$S, $10^{\circ}$ - $40^{\circ}$S, $40^{\circ}$ - $60^{\circ}$S, $60^{\circ}$ - $90^{\circ}$S). In the Martian tropics (between $10^{\circ}$N and $10^{\circ}$S), the eastward-westward alternating pattern associated with the SAO was seen at altitudes between 100 Pa and 1 Pa, and most strongly at pressures above 30 Pa. This region is shown in more detail in Figure \ref{fig:two}. The zonal wind begins with an eastward wind at the start of each MY, changing to a westward wind some time afterwards. This oscillation is evidently repeated twice a year, as in the experiments of \citet{2008GeoRL..3523202K}. In contrast to the cycles found in the uniform dust experiments by \citet{2008GeoRL..3523202K}, however, during each MY one of the two semi-annual cycles appeared to be stronger than the other. Clearly intensified eastward winds were also noticeable between $L_{s}\sim190^{\circ}$ and ${\sim}230^{\circ}$ during the MY 25 global-scale dust storm (GDS). Normally, an eastward wind dominated below 20 Pa, but it was strongly reinforced and extended to higher altitudes (${\sim}$1 Pa) during this strong dust event. 

Apart from the tropics, the zonal wind within other latitude bands across the whole planet was also examined in the current study. It was clear from these that an SAO feature was also evident in latitude bands other than the tropics, though it was less pronounced towards the poles. The SAO pattern was also quite noticeable in the middle atmosphere (100 Pa - 1 Pa) of the sub-tropics ($10^{\circ}$ - $40^{\circ}$N and $10^{\circ}$ - $40^{\circ}$S) with the zonal-mean zonal wind direction oscillating from eastward to westward within a half-year cycle. Similarly to the tropics themselves, the two cycles in each year were of unequal magnitude, such that only one of the two cycles could be seen to extend throughout the whole atmosphere above 200 Pa. When the westward phase of the first SAO cycle dominated in the northern atmosphere, this corresponded to the westward phase of the second SAO cycle in the southern hemisphere extending vertically to the whole atmosphere above ${\sim}$200 Pa. This phenomenon was coincident with the annual cycle of the motion of the sub-solar point in latitude - which is located in the Northern Hemisphere in the first half-year, moving to the Southern Hemisphere in the second half-year. The sub-tropical SAO cycle in each hemisphere thus tends to be stronger when the sunlight is directly imposed upon that hemisphere. The SAO could also be seen in the middle and high latitude bands ($60^{\circ}$ - $90^{\circ}$N, $40^{\circ}$ - $60^{\circ}$N, $40^{\circ}$ - $60^{\circ}$S, $60^{\circ}$ - $90^{\circ}$S) but much less strongly. This did not lead to a full reversal of the wind direction, as in the tropical atmosphere at altitudes higher than 30 Pa, but tended to slow down the prevailing eastward wind during the weaker cycle of the two in a given year.

\subsection{Singular spectrum analysis}

To further study the Martian SAO, a Singular Spectrum Analysis (SSA) was conducted on the daily-averaged zonal-mean zonal wind, in order to isolate the semiannual variations in each latitude band. SSA is a statistical technique applied in the time domain that is widely used in signal processing \citep{1984...pikegeneralised} and is similar to the Empirical Orthogonal Function (EOF) analysis which is usually applied in the spatial domain \citep[e.g.][]{1987MWRv..115.1083B,1991JAtS...48..780G} and decomposes the signal into separate contributions from a set of mutually orthogonal patterns in space and/or time. Compared to other types of spectral analysis, the filters derived in SSA are not prescribed \textit{a priori}, but are determined, optimally with respect to variance, from the data themselves \citep[e.g.][]{1996Icar..120..344C} and therefore focus attention on signal components that contribute most strongly to the signal variance. Further, much more information can be obtained by studying the principal components of particular eigenvectors. SSA was employed here to demonstrate selectively the relative importance of any SAO signal compared to other oscillations, especially in those latitudes and altitudes where the SAO signal was superimposed upon other components.

\begin{figure}[t]
    \centering
    \includegraphics[scale=0.27]{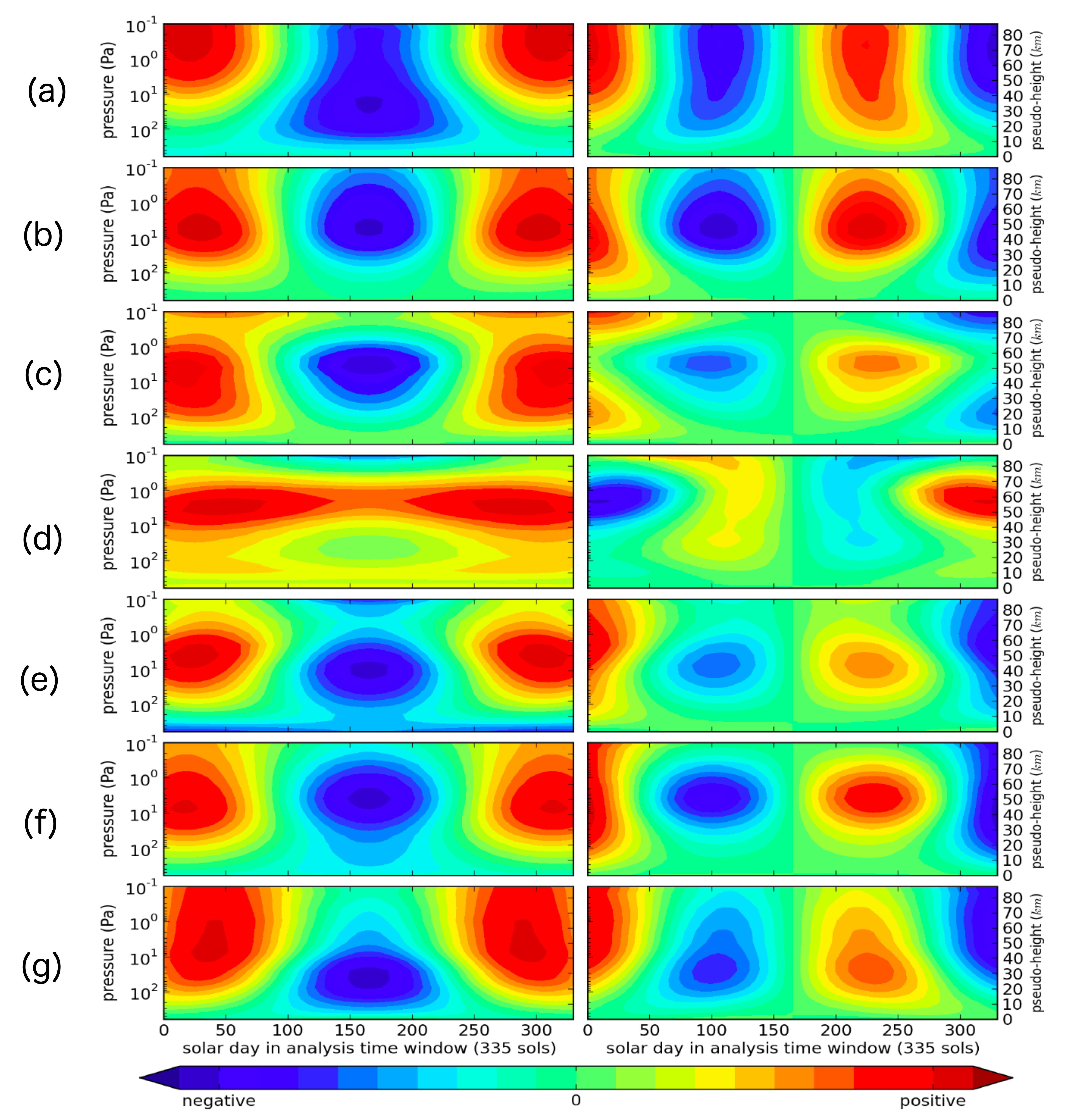}
    \caption{The eigenvectors of the third (left column) and fourth (right column) eigenvalues in different latitude bands, (a) for $60^{\circ}$ - $90^{\circ}$N, (b) for $40^{\circ}$ - $60^{\circ}$N, (c) for $10^{\circ}$ - $40^{\circ}$N, (d) for $10^{\circ}$N - $10^{\circ}$S, (e) for $10^{\circ}$ - $40^{\circ}$S, (f) for $40^{\circ}$ - $60^{\circ}$S, and (g) for $60^{\circ}$ - $90^{\circ}$S. Note that eigenvector 3 is typically dominated by the SAO and eigenvector 4 by its quadri-annual harmonic.}\label{fig:three}
\end{figure}

In order to smooth the day-to-day fluctuations of the zonal wind fields, a 5-sol running-mean was applied to the diurnally-sampled zonal-mean zonal wind. Afterwards, the data was resampled in time by choosing every fifth data point for SSA analysis. The analysis time window here was chosen to be 335 sols (67 resampled data points), corresponding to approximately half a Martian year. The robustness of the analysis was tested by comparing results over at least a 60-sol range ($335\pm30$ sols) in window length to ensure that the results were not sensitive to the choice of this time window in relation to the period of the semiannual oscillation. 

{\renewcommand{\arraystretch}{1.5}
\begin{table}
    \centering 
    \singlespacing
    \begin{tabular}{|c|c|c|c|c|c|c|c|}
        \hline
        latitude band                 & PC 1   & PC 2   & PC 3   & PC 4  & PC 5  & PC 6  & total contribution \\
        \hline 
        $60^{\circ}$ - $90^{\circ}$N  & 42.6\% & 40.2\% & 3.5\%  & 1.9\% & 1.7\% & 1.0\% & 90.9\% \\
        \hline
        $40^{\circ}$ - $60^{\circ}$N  & 40.5\% & 40.3\% & 5.9\%  & 3.6\% & 1.3\% & 1.0\% & 92.6\% \\ 
        \hline
        $10^{\circ}$ - $40^{\circ}$N  & 22.6\% & 22.4\% & 14.5\% & 7.8\% & 4.3\% & 3.2\% & 74.8\% \\   
        \hline     
        $10^{\circ}$S - $10^{\circ}$N & 19.6\% & 16.9\% & 11.1\% & 6.6\% & 4.9\% & 4.7\% & 63.8\% \\  
        \hline      
        $10^{\circ}$ - $40^{\circ}$S  & 39.8\% & 30.7\% & 11.7\% & 4.5\% & 1.4\% & 1.0\% & 89.1\% \\
        \hline
        $40^{\circ}$ - $60^{\circ}$S  & 41.2\% & 34.5\% & 11.1\% & 4.0\% & 1.0\% & 0.9\% & 92.7\% \\
        \hline
        $60^{\circ}$ - $90^{\circ}$S  & 36.6\% & 28.1\% & 10.3\% & 6.0\% & 3.5\% & 1.7\% & 86.2\% \\
        \hline 
    \end{tabular}
    \caption{The contribution of the first 6 eigenvalues to the variance of the final solution of zonal wind.}\label{tab:one}
\end{table}
}

SSA was carried out via the singular value decomposition of a covariance matrix, constructed by evaluating the time-averaged covariances of zonal mean zonal wind with respect to both altitude (pressure) levels and a set of discrete time lags, within each latitude band. The eigenvectors of such an SSA therefore represent two-dimensional recurrent patterns in both altitude (pressure) and time within the time window of the dataset. Eigenvectors are ordered with respect to decreasing eigenvalue, which is proportional to the corresponding with variance fraction ``explained'' by each eigenvector pattern. Some example eigenvectors are illustrated in Figure \ref{fig:three}, which shows the form of eigenvectors 3 and 4 for each of the set of latitude bands considered in this study. This pair of eigenvectors tends to be dominated by the semi-annual signal although, at most latitudes, eigenvector 4 is more strongly dominated by the quadri-annual harmonic of the SAO, and reveals the structure of the semi-annual signal in both height and time of year in each latitude band.

Because the first 6 eigenvalues already capture a large contribution to the variance of the final solution of zonal wind (increasing from a minimum of 63.8\% in the tropics to 92.7\% in the middle and high latitudes; the details can be found in Table \ref{tab:one}), their frequencies (and other characteristics) are discussed here in detail (and are shown in Figure \ref{fig:four}). In the middle ($40^{\circ}$ - $60^{\circ}$N Figure \ref{fig:four}b, $40^{\circ}$ - $60^{\circ}$S Figure \ref{fig:four}f) and high latitude bands ($60^{\circ}$ - $90^{\circ}$N Figure \ref{fig:four}a, $60^{\circ}$ - $90^{\circ}$S Figure \ref{fig:four}g), the first two leading Principal Components (PCs) typically presented a predominantly annual oscillation, while in the sub-tropical bands ($10^{\circ}$ - $40^{\circ}$N Figure \ref{fig:four}c, $10^{\circ}$ - $40^{\circ}$S Figure \ref{fig:four}e) and in the tropics ($10^{\circ}$N - $10^{\circ}$S Figure \ref{fig:four}d), the influence of SAO perturbations was noticeable in the first two PCs, though the main features of the annual oscillation still dominated. This suggests that the annual oscillation component still contributed most to the zonal-mean zonal wind variations at these latitudes.

{
\begin{figure}
    \centering
    \includegraphics[scale=0.84]{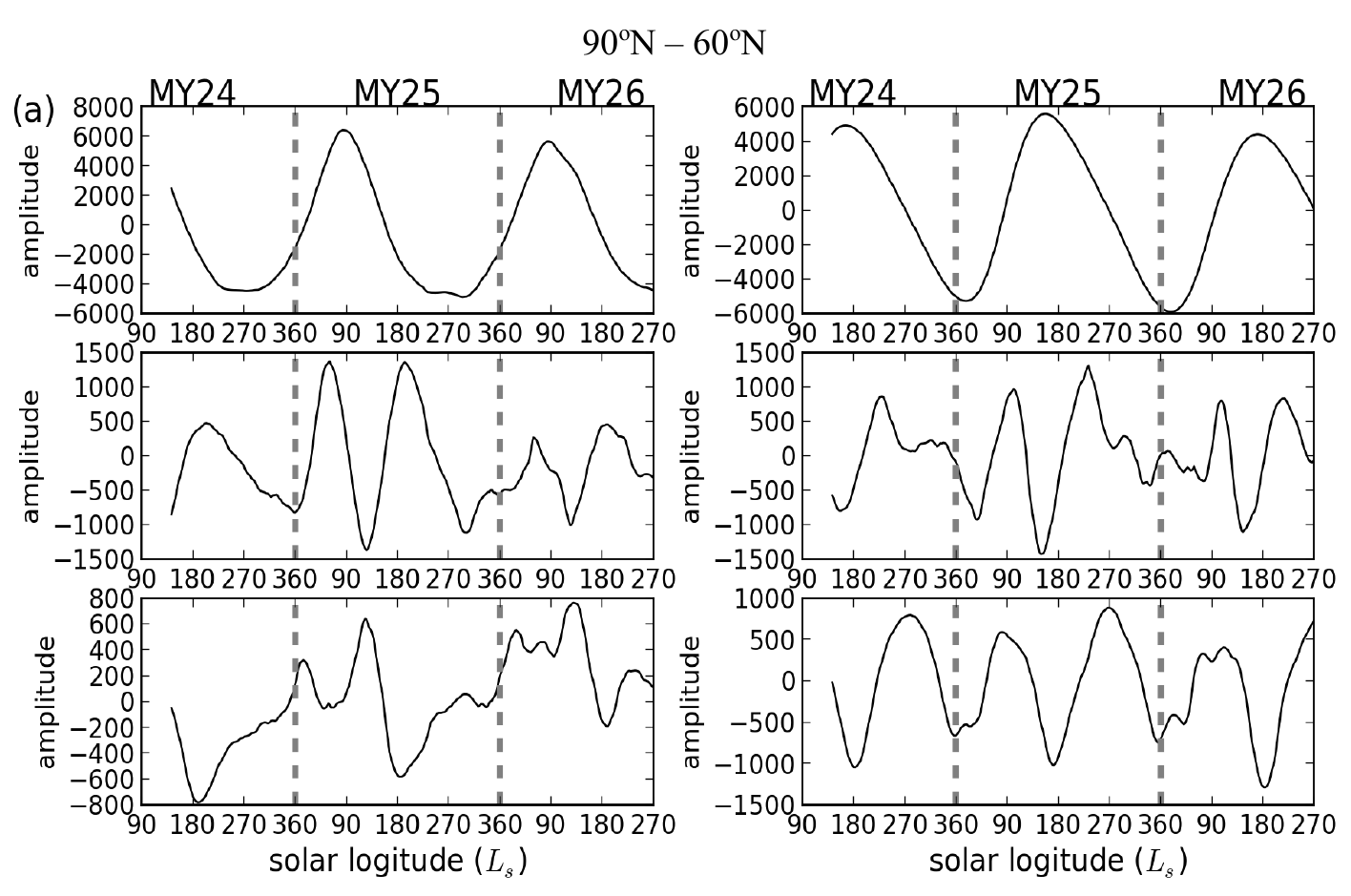}
    \vfill
    \includegraphics[scale=0.84]{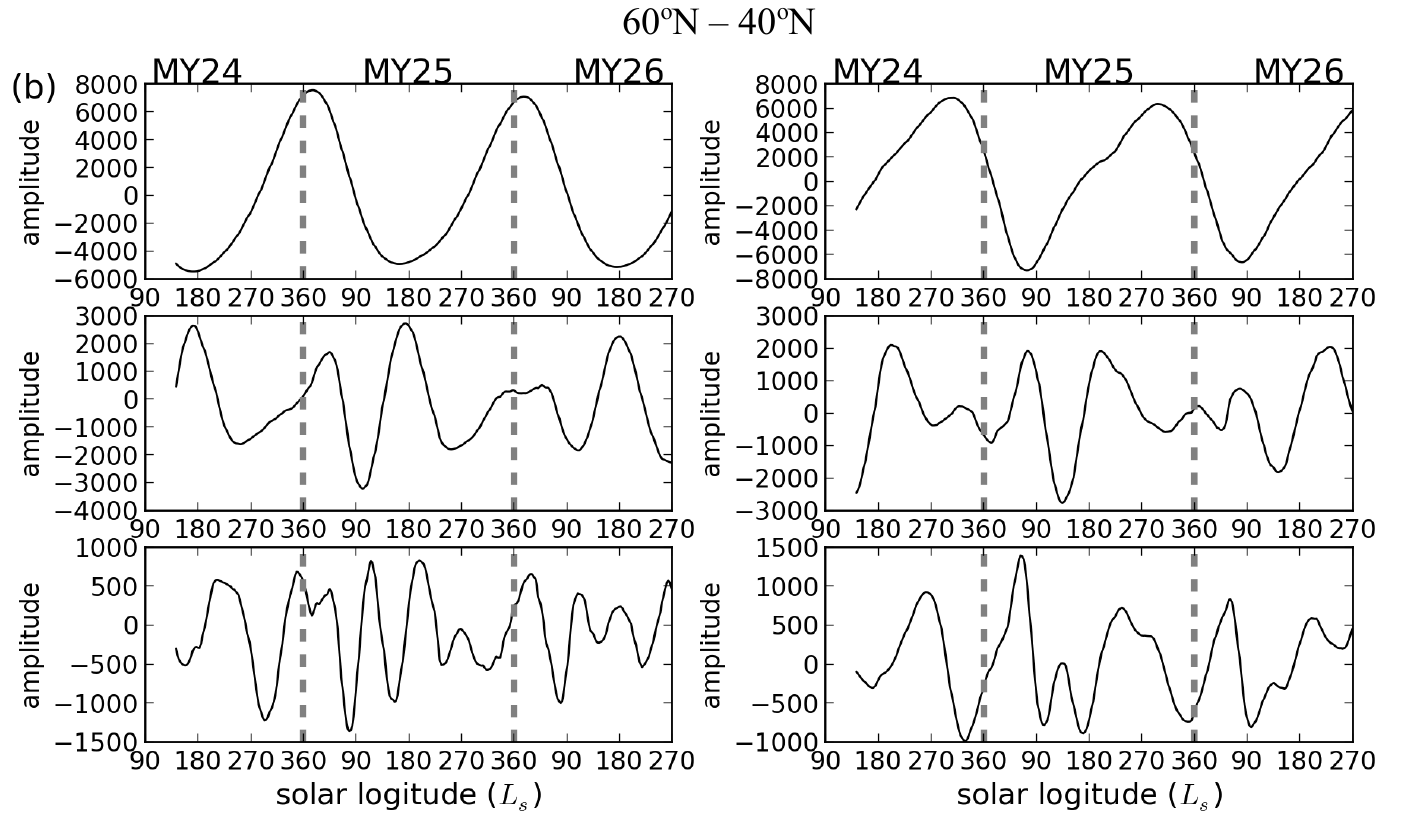}
    \vfill
    \includegraphics[scale=0.84]{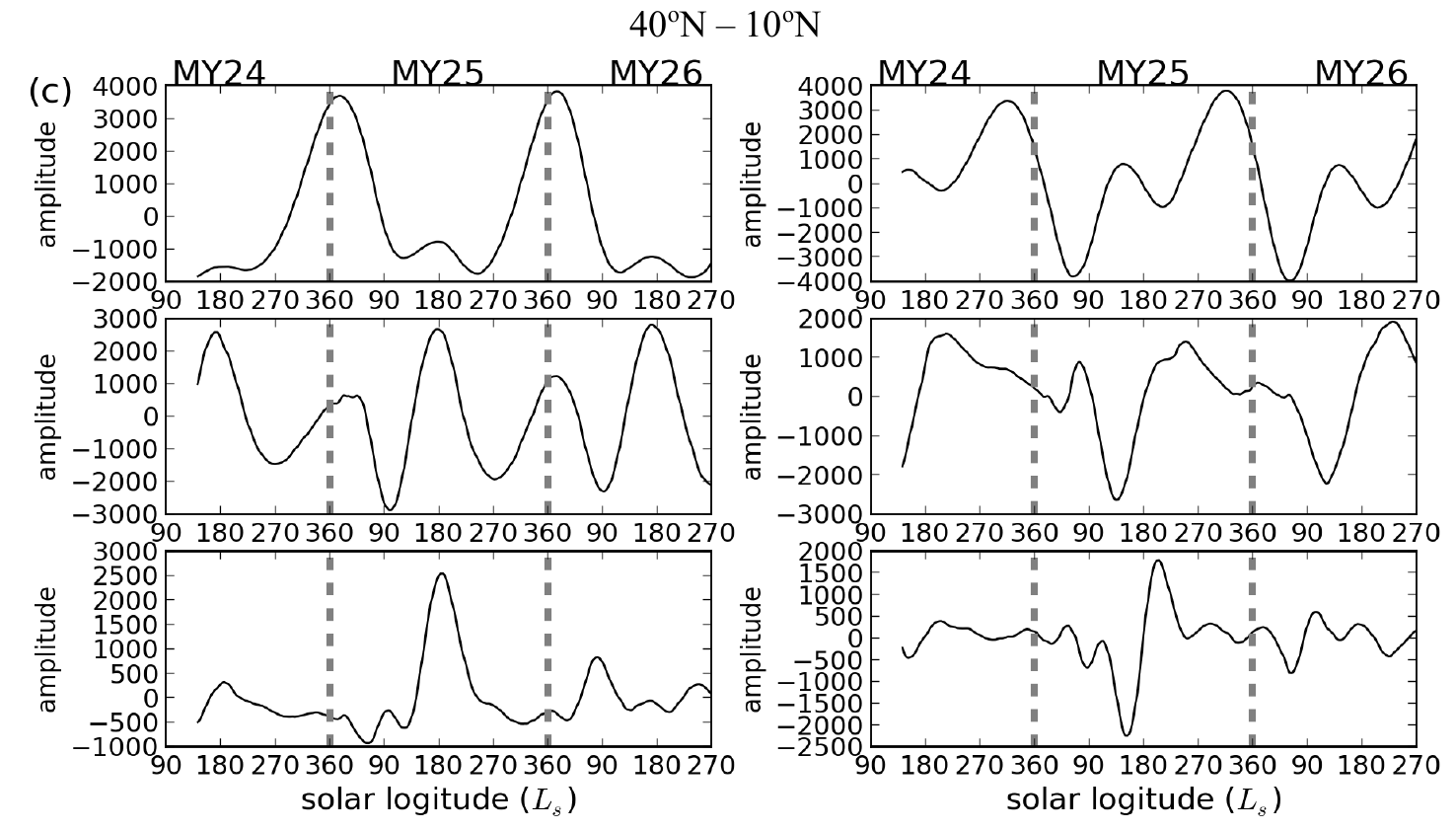}
\end{figure}
\begin{figure}
    \centering
    \includegraphics[scale=0.84]{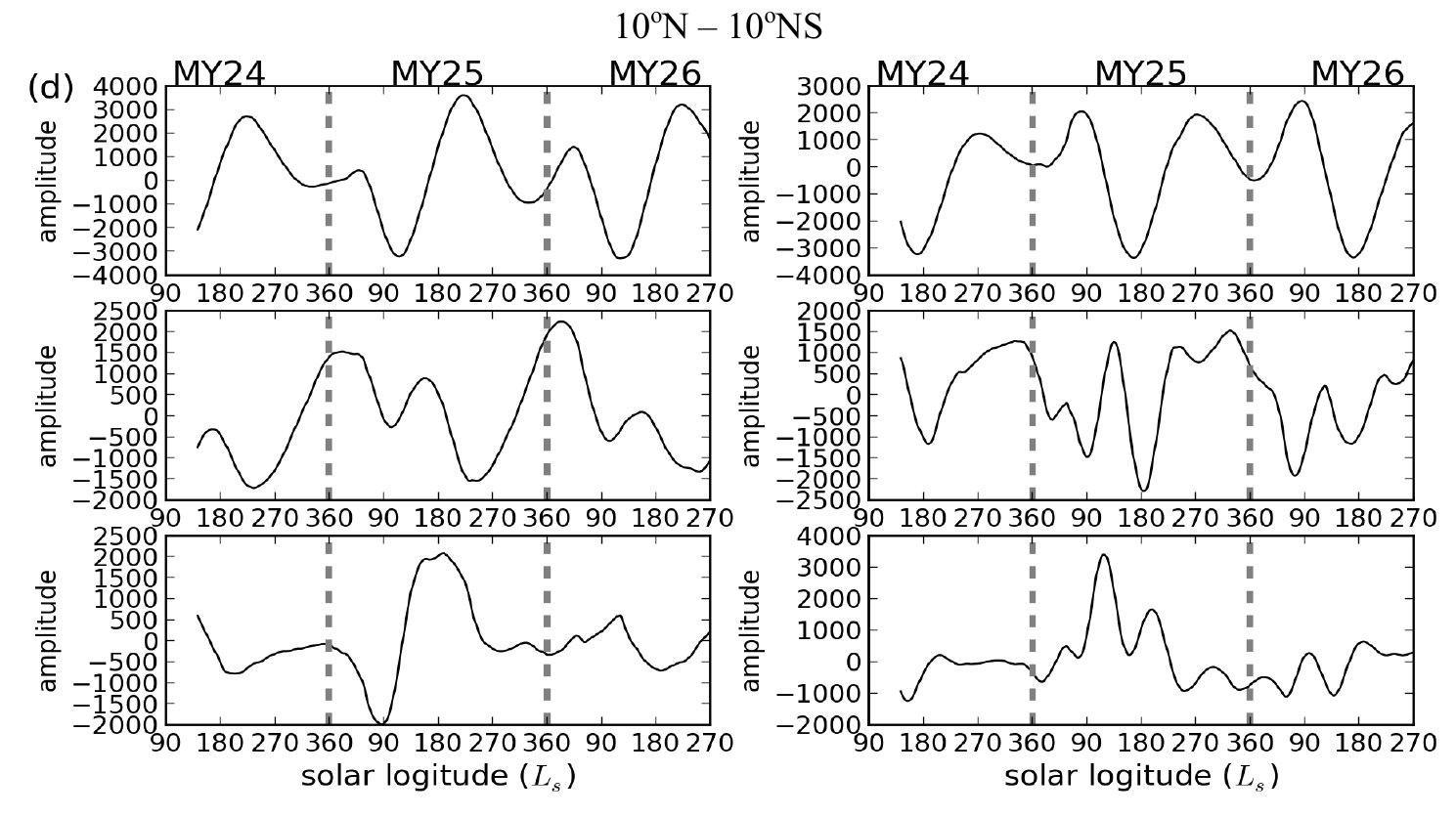}
    \vfill
    \includegraphics[scale=0.84]{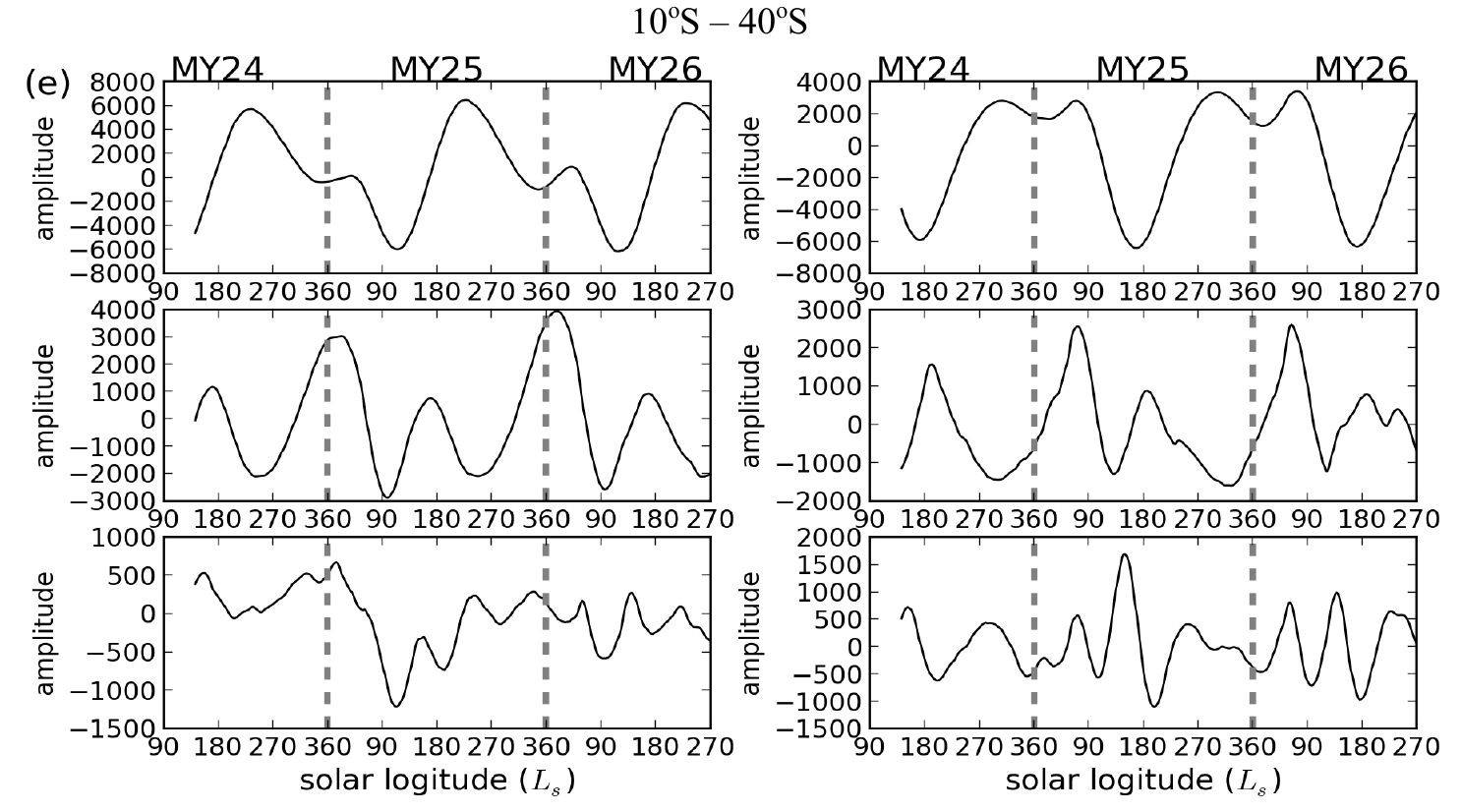}
    \vfill 
    \includegraphics[scale=0.84]{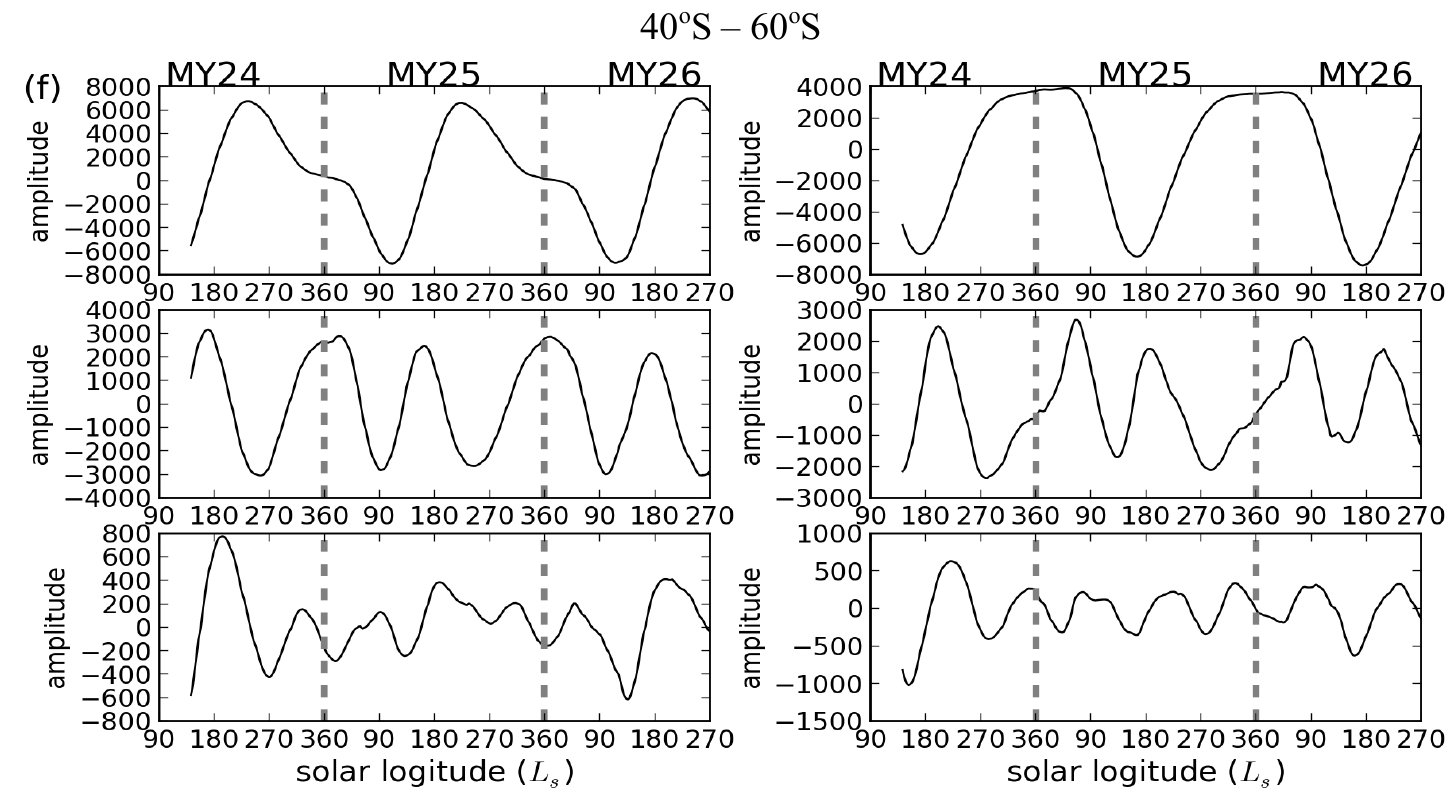}
\end{figure}
\begin{figure}
    \centering
    \includegraphics[scale=0.84]{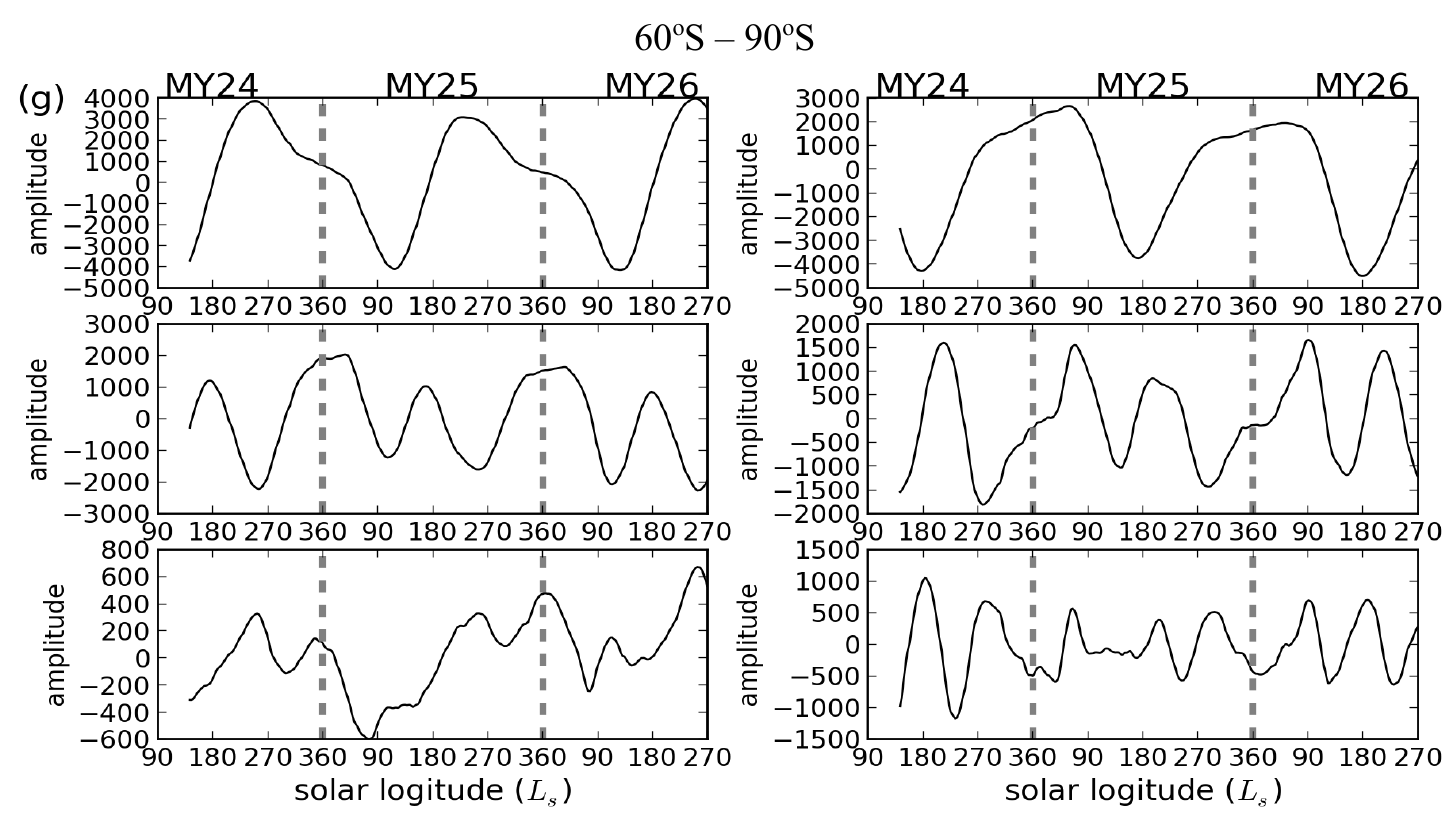}
    \caption{The PCs of the first 6 eigenvalues in different latitude bands. In each group of figures the upper two frames represent the first pair of PCs, the middle two represent the second pair and the lower two the third pair of PCs, (a) for $60^{\circ}$ - $90^{\circ}$N, (b) for $40^{\circ}$ - $60^{\circ}$N, (c) for $10^{\circ}$ - $40^{\circ}$N, (d) for $10^{\circ}$S - $10^{\circ}$N, (e) for $10^{\circ}$ - $40^{\circ}$S, (f) for $40^{\circ}$ - $60^{\circ}$S, and (g) for $60^{\circ}$ - $90^{\circ}$S.}\label{fig:four}
\end{figure}
}

In all the latitude bands, the third and fourth PCs captured most of the SAO signal (all the middle rows of Figure 4). Within each latitude band, five relatively clear cycles could be seen within the whole analysis period. In the Southern Hemisphere including $10^{\circ}$ - $40^{\circ}$S, $40^{\circ}$ - $60^{\circ}$S, $60^{\circ}$ - $90^{\circ}$S, SAO signals were more obvious in both the third and fourth PCs than those in the Northern Hemisphere, and the amplitudes of the signals were more consistent. In the Northern Hemisphere, one of the two PCs in each latitude band exhibited a relatively clear half-yearly oscillation, while the other had some small oscillations superimposed on top of the semiannual cycles, which were presumably due to the impact of interactions on other timescales. 

Except in the high latitudes of the Northern Hemisphere ($60^{\circ}$ - $90^{\circ}$N), the fifth and sixth PCs mainly represent the high-frequency transient patterns, especially in the Southern Hemisphere. The fifth and sixth PCs in the high latitudes of the Northern Hemisphere exhibited signals with similar cycles to the SAO as in the third and fourth PCs. It is interesting to note that the fifth and sixth PCs in the tropics ($10^{\circ}$S - $10^{\circ}$N) and sub-tropics of both hemispheres ($10^{\circ}$ - $40^{\circ}$N, $10^{\circ}$ - $40^{\circ}$S) showed abrupt changes of amplitude during the MY25 GDS (corresponding to mid-2001 on Earth). The further relationship between these two components, corresponding to an increase of actual zonal wind and perhaps connected to the triggering/response and evolution of the MY25 GDS, could be an interesting topic to investigate further in future.

\section{A TEM momentum budget for the SAO in the tropics}\label{sec:tem}

In order to identify the dominant processes generating the Martian SAO, the pressure-coordinate TEM equation for the zonal mean zonal wind \citep[see e.g.][]{1978JAtS...35..175A} was applied to decompose the terms contributing to the acceleration of the zonal-mean flow. If all the terms contributing to the zonal-mean zonal wind tendency were moved to the right-hand side of the equation, this tendency can be written as follows,
\begin{equation}
    \overline{u}_{t}=\left(f-\frac{(\overline{u}\cos\varphi)_{\varphi}}{r_{0}\cos\varphi}\right)\overline{v}^{\ast}
                      - \overline{u}_{p}\overline{\omega}^{\ast} 
                      - \frac{1}{r_{0}\cos^{2}\varphi}\left(S_{(\lambda\varphi)}\cos^{2}\varphi\right)_{\varphi} - 
                      \left(S_{(\lambda p)}\right)_{p}+\overline{X}, \label{eq:tem}
\end{equation}
where $\overline{u}$ is the zonal-mean zonal wind, $f$ is the Coriolis parameter, $\varphi$ is latitude and $r_{0}$ is the radius of the planet, while the residual meridional circulation $\left(0,\overline{{v}}^{\ast},\overline{\omega}^{\ast}\right)$ is defined by 
\begin{equation*}
    \overline{v}^{\ast}=\overline{v}-\left(\frac{\overline{v^{\prime}\theta^{\prime}}}{\overline{\theta}_{p}}\right)_{p}, \quad\quad
    \overline{\omega}^{\ast}=\overline{\omega}+\frac{1}{r_{0}\cos\varphi}\left(\frac{\overline{v^{\prime}\theta^{\prime}}}{\overline{\theta}_{p}}\cos\varphi\right)_{\varphi},
\end{equation*}
and 
\begin{equation*}
    S_{(\lambda\varphi)}=\overline{v^{\prime}u^{\prime}}-\overline{u}_{p}\frac{\overline{v^{\prime}\theta^{\prime}}}{\overline{\theta}_{p}}, \quad\quad 
    S_{(\lambda p)} = \overline{\omega^{\prime}u^{\prime}}+\left(\frac{(\overline{u}\cos\varphi)_{\varphi}}{r_{0}\cos\varphi}-f\right)\frac{\overline{v^{\prime}\theta^{\prime}}}{\overline{\theta}_{p}}.
\end{equation*}
The $\overline{X}$ in (\ref{eq:tem}) represents unspecified zonal components of parameterized forcing or dissipation processes, such as friction or other nonconservative mechanical forcings. 

The terms containing $\overline{v}^{\ast}$ represent the acceleration of the mean zonal wind by meridional advection of momentum, while the second part (containing $\overline{\omega}^{\ast}$) represents the forcing due to vertical momentum advection. The third group of terms results from the non-zonal eddies, and is proportional to the divergence of the Eliassen-Palm flux (hereafter called the total wave forcing). These three contributions were computed respectively from the assimilated meteorological fields from MACDA and averaged in the tropics (between latitudes $10^{\circ}$S and $10^{\circ}$N).

To further investigate the roles of different zonal wavenumber components on the zonal-mean zonal wind, a two-dimensional (i.e. in the temporal and zonal dimensions, respectively) Fast Fourier Transform (FFT) was applied to the MACDA 2-hourly outputs to decompose the meteorological variables into various spatial and temporal harmonics. The contributions from different wave spectral components could be calculated, therefore, through the formulation of the EP flux terms using the decomposed meteorological variables. The different wave components decomposed in the present work consisted of 1) quasi-stationary waves with periods longer than 30 sols, 2) transient waves with periods longer than 1 sol and up to 30 sols, 3) westward thermal tides with periods of 1 sol or shorter, 4) eastward Kelvin waves with periods of 1 sol or shorter.

\begin{figure}[t]
    \centering
    \includegraphics[scale=0.65]{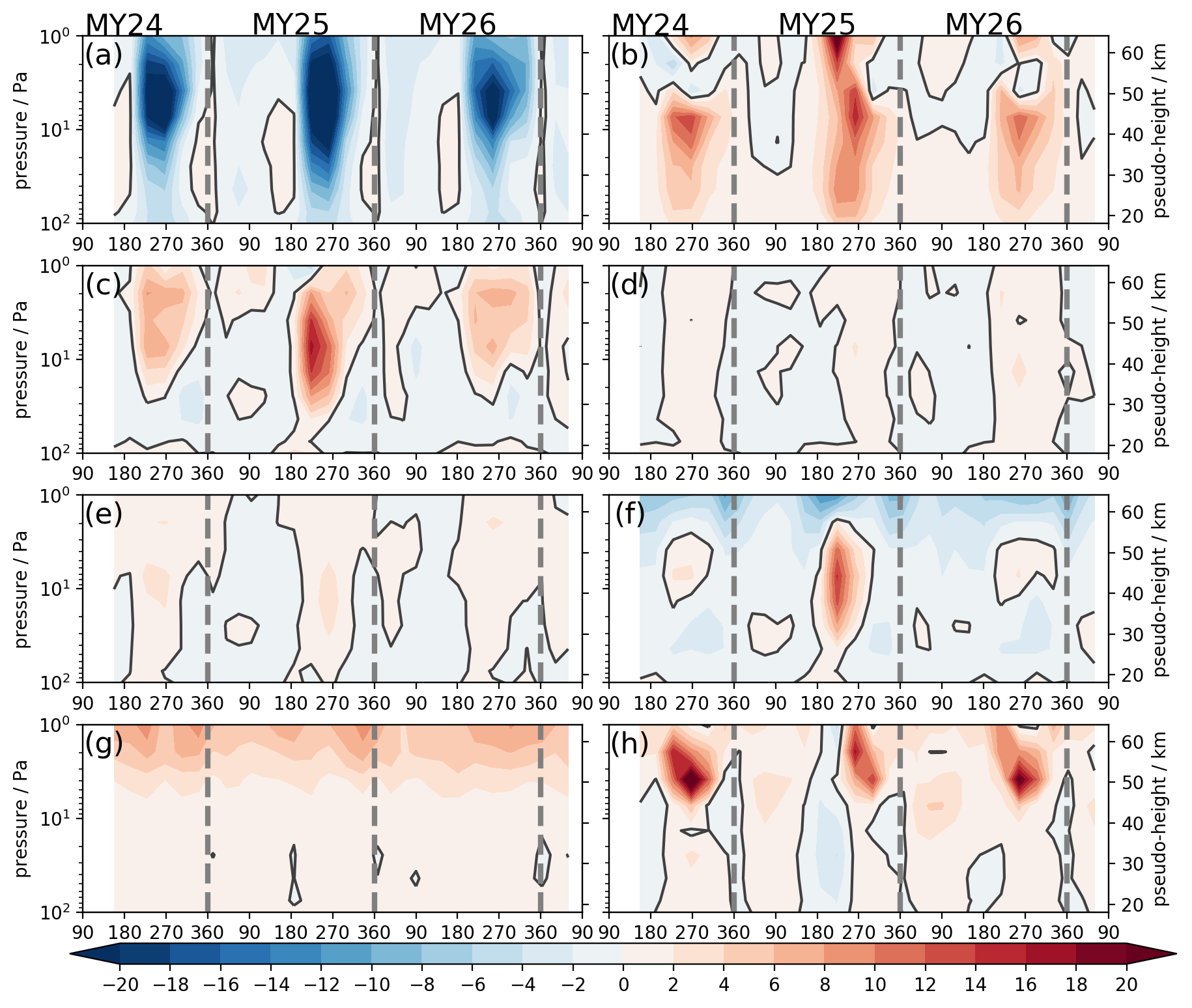}
    \caption{The forcing terms on the zonal-mean zonal wind, decomposed from the TEM zonal momentum equation and averaged, using a 60-sol time window, between latitudes $10^{\circ}$N and $10^{\circ}$S over the same vertical domain as shown in Figure \ref{fig:two}.  The panels show tendencies due to (a) meridional advection, (b) vertical advection, (c) total forcing from different groups of zonal wavenumber-frequency components, (d) quasi-stationary waves only, (e) transient waves only, (f) westward thermal tides only, and (g) eastward Kelvin waves. Panel (h) shows the calculated residual of the actual zonal-mean acceleration calculated from the winds in Figure 2 (also averaged with a 60-sol time window) minus the sum of terms (a), (b) and (c). All terms are in units of m\,s$^{-1}$\,sol$^{-1}$, and only the zero contour line is shown in order to distinguish the eastward (red) and westward (blue) accelerations.}\label{fig:five}
\end{figure}

As the SAO phenomenon mainly happened above 100 Pa and up to altitudes of 1 Pa, the diagnosis of the forces driving it is focused on this part of the Martian atmosphere. Figure \ref{fig:five} shows the decomposed contributions to the torque acting on the zonal-mean zonal wind, $\overline{u}_{t}$. Within this 3-year dataset, the meridional advection (Figure \ref{fig:five}a) provided the strongest westward (retrograde) forcing to the zonal-mean zonal wind, especially around the solstices. In the MY with a GDS (MY 25), the westward forcing due to meridional advection was stronger during northern fall and winter, with the contours of westward forcing $\geq$ 10 m\,s$^{-1}$\,sol$^{-1}$ penetrating lower down (to $>$ 30 Pa) than in other MYs. The maximum of westward forcing was ${\sim}$ 20 m\,s$^{-1}$\,sol$^{-1}$ for the average of 60-sol time windows. The meridional advection forcing showed no such substantial difference in magnitude for both solstices in a year without a global dust storm. The meridional advection switched to eastward (prograde) forcing at the equinoxes, but was not strong enough on its own to completely reverse the direction of the zonal-mean zonal wind from westward to eastward. This eastward forcing indeed contributed to the eastward zonal-mean zonal wind between about 3 Pa and 70 Pa at the equinoxes. Overall, the meridional advection displays a clear SAO cycle correlated with the variation of zonal-mean zonal wind.

In contrast, the vertical advection term (Figure \ref{fig:five}b) mainly supplies eastward forcing to the zonal-mean zonal wind, but this forcing was significantly stronger during the northern fall and winter seasons (maximum ${\sim}$ 11 m\,s$^{-1}$\,sol$^{-1}$ in 60-sol average at altitudes below 10 Pa). Mild westward forcing could mainly be seen between 15 Pa and 3 Pa during northern spring and summer with a magnitude no larger than 1 m\,s$^{-1}$\,sol$^{-1}$.

The total forcing from different non-zonally symmetric waves (Figure \ref{fig:five}c) exhibited more variation than the other two terms in (\ref{eq:tem}). In MY 25 with a GDS, wave activity only provided westward forcing around 10 Pa in the first half of the year, while in MY 26 without a GDS, it also provided eastward forcing during $L_{s}\approx30^{\circ}\sim70^{\circ}$. Apart from this, the total wave forcing exhibited a half-year cycle above ${\sim}$ 40 Pa. The phase of this oscillation was almost opposite to the forcing due to meridional advection above ${\sim}$ 3 Pa. Below 3 Pa altitude, the total wave forcing also partly contributed some westward forcing to the SAO of the zonal-mean zonal wind. The eastward forcing below 3 Pa started at $L_{s} \sim 180^{\circ}$, but it expanded throughout most of the second SAO. As a result, the eastward forcing due to total wave forcing appeared to support the eastward component of the SAO, but also weakened the westward forcing to some degree. Compared to the study by \citet{2008GeoRL..3523202K}, in the present analysis the eastward forcings in the second half of the year were obviously stronger in all three terms (meridional advection, vertical advection and total wave forcing). This is likely because of the increased diabatic forcing due to this being a more dusty season in the second half of the year (perihelion) than assumed by \citet{2008GeoRL..3523202K} in their prescribed, smooth dust distribution. This effect was mentioned by \citet{2008GeoRL..3523202K} as well, when they discussed the forcing due to meridional advection. Also, the eastward forcing in the vertical advection and total wave forcing terms reached their maximum around $L_{s} \sim 220^{\circ}$ in the current study instead of $\sim 180^{\circ}$ in the study by \citet{2008GeoRL..3523202K}. This might be because the dustiest period in general does not usually happen at equinox, and this was not represented by a seasonally uniform dust distribution in a free-running simulation as was the case in the study by \citet{2008GeoRL..3523202K}. 

Among different wave forcings, quasi-stationary waves (Figure \ref{fig:five}d) mainly supplied weak westward momentum to the zonal-mean zonal wind in the first half of the year, and then supplied eastward momentum in the second half of the year. In the study of \citet{2008GeoRL..3523202K}, the synoptic transient waves were claimed to have little contribution to the tendency of zonal-mean zonal wind. In our current study, in contrast, the transient waves (Figure \ref{fig:five}e) exhibited a similar effect to that of quasi-stationary waves. Such a role for baroclinic transient eddies at low latitudes is consistent with the recent results of \citet{2018Icar..307..150H}, who found instances in observations where northern circumpolar baroclinic disturbances propagated along western boundary currents adjacent to topography towards the tropics, to be detected by the REMS instrument on the Curiosity Rover. The apparent peak in prograde zonal forcing by  baroclinic transients at levels around 10-20 Pa is also consistent with the peak in rms temperature fluctuations due to baroclinic waves by \citet{2016Icar..264..456L}. \citet{2016Icar..276....1B} also found that baroclinic transients during the GDS in MY25 were enhanced at altitudes above 200 Pa and relatively suppressed nearer the surface, consistent with the present work.

The westward thermal tides (Figure \ref{fig:five}f) displayed an SAO cycle above 10 Pa but with opposite forcing tending to counteract the momentum supplied by the meridional circulation. In the first half of the year, the thermal tides imposed a westward forcing on the zonal-mean zonal wind. But in the second half of the year, its effect varied with pressure. Between 50 Pa and 10 Pa of a typical MY without a GDS (MY 24 and MY 26 in MACDA), the thermal tides provided westward forcing that could contribute to the westward component seen in the second SAO cycle of the zonal-mean zonal wind. Between 10 Pa and 2 Pa or below 50 Pa, the thermal tides provided eastward forcing. We note that they have also been found to lead to super-rotation within the height range of tidal forcing in the Martian atmosphere, given sufficiently dusty conditions \citep{1973GApFD...5..211F,2003JGRE..108.5034L}. The eastward Kelvin waves (Figure \ref{fig:five}g) consistently provided eastward momentum to $u$ above 30 Pa. This is consistent with the study conducted by \citet{2008GeoRL..3523805M}, in which the eastward propagating Kelvin waves were found to lead to strong eastward acceleration in the upper atmosphere associated with their dissipation. On the contrary, the westward forcing yielded by the eastward Kelvin waves in the study by \citet{2008GeoRL..3523202K} is contradictory to the findings in this and other studies \citep[e.g.][]{2008GeoRL..3523805M}. A possible reason for this might be that the Kelvin waves in their model are forced non-physically, rather than generated spontaneously by the model physics. Below 30 Pa, the oscillation of the forcing due to eastward Kelvin waves in our analysis had a weak SAO signature, but with the opposite sign compared to the forcing due to meridional advection. 

It is also worth mentioning the changes of the wave forcing during the GDS year, i.e. MY 25. As generally the thermal tides dominated in the total wave forcing during the GDS, the significant increase of total wave forcing was mainly the result of increasing forcing due to amplified thermal tides. Although the forcing due to quasi-stationary waves and transient waves exhibited some similarity in the annual pattern, they showed different responses to the GDS. The forcing due to transient waves supplied stronger eastward forcing during the GDS event, while the forcing due to quasi-stationary waves only led to a mild increase in the eastward forcing. However, in the year following the MY 25 GDS, the eastward forcing due to quasi-stationary waves was clearly stronger than it was in other MYs. In contrast, the GDS event had almost no impact on the eastward Kelvin waves above 100 Pa.

Finally, the residual of the actual zonal mean wind acceleration minus the sum of the zonal mean meridional advection, vertical advection and wave terms is shown in Figure \ref{fig:five}h. This is smaller than each of the individual terms in Figures \ref{fig:five}a-c below about 5 Pa (45 km), showing that the semi-annual pattern to the acceleration is largely explained by them. It is worth noting, however, that it is still significant in magnitude compared to the actual zonal wind accelerations (when averaged with a 60-sol window) and that in places, especially above 5 Pa, it is over an order of magnitude greater. This is likely to be a result of the strength of diabatic terms in the Martian atmosphere, in particular the short radiative timescale (about 1 sol) may lead to radiation acting as a damping in the residual term and making it more significant than in the case of the Earth. We also note the possibility that the larger residuals seen above 5 Pa may be, at least in part, an artifact of the free response of the model to the assimilation increments which cut off just below this level. Assimilation increments are only applied below about 40 km where TES retrievals exist, but we do not see any clear reduction in the size of the residuals when data is not assimilated and, in general, all the terms shown in Figure \ref{fig:five}a-g are as large during these periods.

\section{Disussion}

In the current study, the phenomenology of the SAO in the zonal-mean zonal wind has been presented and reviewed. Unlike the preceding study of a Martian SAO by \citet{2008GeoRL..3523202K}, the present study used a dataset from a model directly constrained by observations to follow the actual weather during three Mars Years \citep{2011..MACDA,2014GSDJ....1..129M,2007Icar..192..327L}. In the analysis, the SAO phenomenon appeared to happen not only in the tropics, but also extended its influence to higher latitudes. This oscillation appeared not always to manifest itself in completely reversing the direction of the zonal-mean zonal wind, but at some latitudes or pressure levels it was found to appear as the deceleration or acceleration of a prevailing unidirectional zonal-mean zonal wind. A Singular Systems Analysis of $u(t,p)$ enabled the SAO component of atmospheric variability to be efficiently isolated and analysed, demonstrating that this component plays a significant role in the seasonal variability of the Martian atmosphere. It was evident that the second pair of principal components (PC) was mainly dominated by this SAO cycle, though the SAO signal was also strong enough to appear in the first pair of eigenvectors and corresponding PCs.

Through analyzing the different terms in TEM equation, the nature of the forcing processes driving the tendency of the zonal-mean zonal wind in the tropics was studied. The meridional advection term, associated with seasonal variations in the thermally-direct meridional circulation at low latitudes, exhibited a clear SAO cycle strongly correlating with the observed oscillation of zonal-mean zonal wind, and supplied the majority of the westward (retrograde) forcing in the SAO cycle. The vertical advection term mainly supplied eastward forcing to the zonal-mean zonal wind with a strong annual cycle. The decomposition into different wave contributions suggested that the quasi-stationary waves and transient waves partly supplied both eastward and westward forcings to the first cycle of SAO at different phases, but they tended to weaken the strength of the westward component in the second SAO of the year. The thermal tides evidently dominated in the pattern of total wave forcing. They mainly supplied the westward forcing in the first half of the year, and also supplied the westward forcing for the second cycle of the year between 50 Pa and 10 Pa. In the total wave forcing, the mild eastward forcing above 3 Pa in the first half of year was found to be mainly a result of eastward Kelvin waves. 

The diagnostic results here based on MACDA were somewhat different from the work of \citet{2008GeoRL..3523202K}, that was based on a free-running model simulation with a uniform assumed dust distribution. The observed dust distribution \citep{2015Icar..251...65M} is clearly not consistent with such an assumption, so we should not be unduly surprised that this might lead to discrepancies between models and observations. The processes driving the tendency of zonal-mean zonal wind in the study of \citet{2008GeoRL..3523202K} were mainly different from ours in the following ways: (1) the forcing due to the meridional advection in our study was less asymmetric between the two SAO cycles of a year, (2) the total wave forcing and decomposed thermal tides did not consistently display the oscillation as an SAO of zonal-mean zonal wind in every pressure level, (3) the strength of the forcing due to transient waves was comparable to that of the quasi-stationary waves, and (4) similar to the Earth, eastward Kelvin waves mainly supplied eastward momentum to the zonal-mean zonal wind, consistent with their dissipation of wave action. Finally, (5) the meridional advection and quasi-stationary wave forcing in the simulations of \citet{2008GeoRL..3523202K} exhibited a peak or trough close to the surface which was not seen in the present work. The reason for this is not clear and should be investigated further in future work.

As the period of this dataset also included the period of the MY 25 GDS, the diagnosis also highlighted some distinctive features of the forces driving the circulation in this GDS year. The meridional advection supplied weaker westward momentum before the onset of MY 25 GDS, and the eastward momentum and westward momentum evident during the GDS were clearly stronger than those in other MYs. Along with the onset of the MY 25 GDS, the vertical advection and total wave forcing were also stronger. Among different wave components, the thermal tides were significantly intensified, most probably due to the enhanced solar heating effect due to increased dust during the GDS \citep[see e.g.][]{2005AdSpR..36.2162L,2014JGRE..119..506G}. Although the forcings due to quasi-stationary waves and transient waves in general displayed a similar pattern, the eastward forcing due to quasi-stationary waves appeared to be stronger in the year after the MY 25 GDS instead of during the MY 25 GDS itself. 

In future work, the TEM analysis should be extended to the latitude bands outside of the tropics, making it possible to discover differences among the different processes across different latitudes. To establish any further link between the features of different forcings and the trigger/response of a GDS, it would be desirable to investigate at least one other GDS event. This will require the extension of reanalysis datasets, such as MACDA, to include MYs with other GDSs, such as occurred in MY 28.

\small 

\section*{Data availability}

The MACDA dataset on which this paper is based can be accessed via the website of the Centre for Environmental Data Analysis of the UK Natural Environment Research Council (\url{https://catalogue.ceda.ac.uk/uuid/01c44fb05fbd6e428efbd57969a11177}) and is documented by \citet{2014GSDJ....1..129M}.

\section*{Acknowledgements}

The authors are grateful to Prof. David Andrews for his insights into the discussion of the formulation of Transformed Eulerian Mean equations. T. Ruan is also grateful to Prof. Philip Stier and Dr. John Wilson for useful discussions concerning this work. L. Montabone acknowledges funding for Mars data assimilation from the National Aeronautics and Space Administration (NASA) under grant no. NNX13AK02G issued through the Mars Data Analysis Program 2012. S. Lewis thanks STFC for funding under grant ST/L000776/1 and the EU under the Horizon2020 programme UPWARDS ref. 633127, while P. Read and N. Lewis acknowledge STFC support under grants ST/I001948/1 and ST/S505638/1. We are grateful to Dr J Battalio and an anonymous referee for their constructive comments on an earlier version of this paper. 

\section*{References}


\end{document}